\documentclass[tighten,times,twocolumn]{aastex63}
\usepackage{multirow}
\usepackage{amsmath}
\usepackage{mathtools}
\usepackage{amssymb}
\usepackage[T1]{fontenc}
\let\tablenum\relax
\usepackage[multi-part-units=single,separate-uncertainty=true]{siunitx}
\newcommand{\ms}[2]{\ensuremath{#1_{\textnormal{#2}}}}
\newcommand{\av}[2]{\ensuremath{\langle#1\rangle}_{#2}}

\newcommand{\Mdot}{\ensuremath{\dot{M}}}
\newcommand{\MdotEdd}{\ensuremath{\ms{\dot{M}}{Edd}}}
\newcommand{\mdot}{\ensuremath{\dot{m}}}
\newcommand{\MSun}{\ensuremath{M_{\sun}}}
\newcommand{\rmodellabel}{R}
\newcommand{\sgra}{Sgr A$^\ast$}
\newcommand{\mes}{M87$^\ast$}
\newcommand{\spin}{\ensuremath{a_{\ast}}}

\newcommand{\radgrmhd}{RadGRMHD}
\newcommand{\xray}{X-ray}
\newcommand{\xrays}{X-rays}

\newcommand{\code}[1]{\texttt{#1}}

\newcommand{\coolingrate}[1]{\ensuremath{\Lambda^{\textnormal{#1}}}}
\newcommand{\F}[1]{\ensuremath{F_{\textnormal{#1}}\left(\thte\right)}}
\newcommand{\gaunt}[1]{\ensuremath{\bar{g}_{\textnormal{ff}}^{\textnormal{#1}}}}
\newcommand{\invu}{\frac{\ms{k}{B}\ms{T}{e}}{h\nu}}
\newcommand{\jnu}[1]{\ensuremath{j_{\nu}}^{\textnormal{#1}}}

\newcommand{\kb}{\ensuremath{\ms{k}{B}}}
\newcommand{\thte}{\ms{\Theta}{e}}
\newcommand{\Lum}[1]{\ensuremath{L^{\textnormal{#1}}}}
\newcommand{\nuLnu}{\ensuremath{\nu L_{\nu}}}
\newcommand{\tpte}{\ensuremath{\ms{T}{p}/\ms{T}{e}}}

\newcommand{\lp}{\left(}
\newcommand{\rp}{\right)}
\newcommand{\ls}{\left[}
\newcommand{\rs}{\right]}

\newcommand{\lb}{\left\{}
\newcommand{\rb}{\right\}}

\newcommand{\spectrumgridsize}{0.5\textwidth}

\DeclareSIUnit{\erg}{erg}
\DeclareSIUnit{\year}{yr}
\DeclareSIUnit{\parsec}{pc}
\DeclareSIUnit{\arcsec}{as}
\DeclareSIUnit{\jansky}{Jy}
\DeclareSIUnit{\solarmass}{\MSun}
\DeclareSIUnit{\astronomicalunit}{au}
\DeclareSIUnit{\pixel}{pixel}
\usepackage[final]{microtype}

\renewcommand{\av}[2]{\ensuremath{\overline{#1}}}
\renewcommand{\av}[2]{\ensuremath{\langle #1\rangle}}

\shortauthors{Yarza et al.}
\begin{document}
\title{Bremsstrahlung \replaced{from radiation}{in} GRMHD models of accreting black holes}

\correspondingauthor{Ricardo Yarza}
\email{ricardo9@illinois.edu}
\author[0000-0003-0381-1039]{Ricardo Yarza}
\affil{Department of Physics, University of Illinois, 1110 West Green Street, Urbana, IL 61801, USA}
\affil{Department of Astronomy, University of Illinois, 1002 West Green Street, Urbana, IL 61801, USA}
\author[0000-0001-6952-2147]{George N. Wong}
\affil{Department of Physics, University of Illinois, 1110 West Green Street, Urbana, IL 61801, USA}
\author[0000-0001-8939-4461]{Benjamin R. Ryan}
\affil{CCS-2, Los Alamos National Laboratory, P.O. Box 1663, Los Alamos, NM 87545, USA}
\author[0000-0001-7451-8935]{Charles F. Gammie}
\affil{Department of Physics, University of Illinois, 1110 West Green Street, Urbana, IL 61801, USA}
\affil{Department of Astronomy, University of Illinois, 1002 West Green Street, Urbana, IL 61801, USA}

\begin{abstract}
\replaced{We use the radiative transfer code \code{grmonty} to compute spectral energy distributions (SED) of accretion disks around slowly accreting supermassive black holes. These SEDs include synchrotron radiation, inverse Compton scattering, and bremsstrahlung.}{The role of bremsstrahlung in the emission from hot accretion flows around slowly accreting supermassive black holes is not thoroughly understood. In order to appraise the importance of bremsstrahlung relative to other radiative processes, we compute spectral energy distributions (SEDs) of accretion disks around slowly accreting supermassive black holes including synchrotron radiation, inverse Compton scattering, and bremsstrahlung.} We compute SEDs for (i) four axisymmetric radiative general relativistic magnetohydrodynamics (RadGRMHD) simulations of $10^{8}M_{\sun}$ black holes with accretion rates between $10^{-8}\dot{M}_{\text{Edd}}$ and $10^{-5}\dot{M}_{\text{Edd}}$, (ii) four axisymmetric RadGRMHD simulations of M87$^\ast$ with varying dimensionless spin $a_\ast$ and black hole mass, and (iii) a 3D GRMHD simulation scaled for Sgr A$^\ast$. At $10^{-8}\dot{M}_{\text{Edd}}$, most of the luminosity is synchrotron radiation, while at $10^{-5}\dot{M}_{\text{Edd}}$ the three radiative processes have similar luminosities. In most models, bremsstrahlung dominates the SED near $\SI{512}{\kilo\electronvolt}$. In the M87$^\ast$ models, bremsstrahlung dominates this part of the SED if $a_{\ast} = 0.5$, but inverse Compton scattering dominates if $a_{\ast}= 0.9375$. Since scattering is more variable than bremsstrahlung, this result suggests that $\SI{512}{\kilo\electronvolt}$ variability could be a diagnostic of black hole spin. In the appendix, we compare some bremsstrahlung formul\ae{} found in the literature.
\end{abstract}


\section{Introduction}\label{sec:intro}
Black holes with dimensionless accretion rate\footnote{$\MdotEdd\equiv4\pi GM\ms{m}{p}/\eta\ms{\sigma}{T}c$, where $M \equiv$ black hole mass, $\ms{m}{p} \equiv$ proton mass, $\ms{\sigma}{T} \equiv$ Thompson cross section, and $\eta \equiv$ nominal efficiency, conventionally taken to be $0.1$} $\mdot\equiv\Mdot/\MdotEdd\leq10^{-3}$ are thought to accrete through optically thin, geometrically thick, radiatively inefficient accretion flows \citep[for a review, see][]{Yuan2014}. The dominant emission processes in these flows are synchrotron radiation and bremsstrahlung, and photons emitted \replaced{due to}{by} these processes may then be upscattered by inverse Compton scattering. Each of these processes has been extensively considered in the accretion flow literature \citep[e.g.,][]{Mahadevan1997,Narayan1995,Esin1996}, but bremsstrahlung is sometimes neglected because synchrotron radiation is energetically dominant at low accretion rates.

The effect of radiative processes on flow dynamics depends on $\mdot$. At very low $\mdot$, \replaced{radiation removes only a negligible fraction of the internal energy before the plasma accretes onto the central black hole}{the plasma radiates only a negligible fraction of its internal energy before being accreted by the central black hole}. Using numerical simulations, \cite{Dibi2012} found that radiative processes are negligible to the flow dynamics if $\mdot\leq 10^{-7}$. Another study \citep{Ryan2017} found this constraint to be $\mdot\leq10^{-6}$ instead, and attributed the difference to their electron temperature prescription \citep{Ressler2015}. As $\mdot$ increases,\deleted{ however,} radiative cooling becomes increasingly important, motivating relativistic radiation magnetohydrodynamics (\radgrmhd{}) models of the flow. As we shall see, bremsstrahlung becomes increasingly \added{energetically} important compared to synchrotron radiation as \mdot{} increases.

Even when the accretion rate is sufficiently low that radiative cooling is unimportant, photons produced at $h\nu \gtrsim \ms{m}{e}c^2$ may still dominate electron-positron pair production \citetext{\citealp{Moscibrodzka2011}; Wong et al. 2020, in preparation}. Pairs created by background photon collisions can influence the structure of the accretion flow in regions where the native plasma density is too low to screen the electric field in the plasma frame, i.e., when the density is less than the \cite{Goldreich1969} density. \replaced{Thus, even when it is energetically negligible, because bremsstrahlung may still dominate the \xray{} component of the radiation field, and affect observables through the pair production mechanism.}{Therefore, even when bremsstrahlung is energetically negligible, it may still dominate the \xray{} and gamma ray radiation field and affect observables via pair production.}
Accurately accounting for the bremsstrahlung component of the radiation field may be important in future particle kinetics simulations \citep{Ford2018,Parfrey2019} and magnetohydrodynamic models of pair production.

The structure of low-$\mdot$ black hole accretion flows is of particular interest with the advent of resolved mm-wavelength images of the black hole at the center of the elliptical galaxy M87
\citep{EHT2019P1} (the black hole is hereafter referred to as \mes{}). \xray{} observations of \mes{} are numerous \citep{Bohringer2001, Wilson2002, DiMatteo2003,Prieto2016,EHT2019P5}. Measuring \mes{}'s mass from gas dynamics yields $m\equiv M/\MSun=3.5\times10^{9}$ \citep{Walsh2013}, from stellar dynamics $m=6.6\times10^{9}$ \citep{Gebhardt2011}, and from interferometric measurements $m=6.5\times10^{9}$ \citep{EHT2019P6}. The distance to \mes{} is approximately $D=\SI{16.8}{\mega\parsec}$ \citep{Blakeslee2009,Bird2010,Cantiello2018}. Estimates and analysis of simulations show that $\mdot\sim10^{-5}$ and that the dimensionless spin ($\spin\equiv J c / G M^2$, where $J$ is the angular momentum of the black hole) satisfies $\lvert\spin\rvert\gtrsim0.5$ \citep[e.g.,][]{EHT2019P5}.

Similarly, resolved mm-wavelength images of \sgra{} are expected in the near future. \sgra{}'s quiescent state has also been observed extensively in the \xray{} \citep{Baganoff2003, Belanger2006,Nowak2012}. Stellar orbit observations suggest $m=4.05\times10^{6}$ \citep{Boehle2016} and $D=\SI{8.18}{\kilo\parsec}$ \citep{Gravity2019}. Analysis of polarized radiation at $\lambda=\SI{1.3}{\milli\meter}$ \citep{Marrone2006} suggests that the accretion rate at $r=20GM/c^{2}$ is in the range $2\times10^{-9}\leq\mdot\leq2\times10^{-7}$, and some GRMHD simulations \citep[e.g.,][]{Moscibrodzka2009} find that $\mdot\sim10^{-8}$ reproduces the observed $\SI{230}{\giga\hertz}$ flux \citep{Marrone2006}. At such low accretion rates, radiative cooling is negligible.

These considerations motivate a study of bremsstrahlung in low-$\mdot$ black holes. The recent development of \radgrmhd{} codes \citep[e.g.,][]{Sadowski2014,Ryan2015} enables self-consistent studies of radiative cooling at accretion rates up to $\mdot\sim10^{-5}$, which permits a more accurate evaluation of the importance of bremsstrahlung.

In this paper, we consider eight \radgrmhd{} models of accreting black holes and one nonradiative GRMHD model for \sgra{} (see Table~\ref{tab:models} for a summary of the simulations). We compute their spectral energy distributions (SEDs) using a modified version of the radiative transfer code \code{grmonty} \citep{Dolence2009} that includes synchrotron radiation, inverse Compton scattering, and bremsstrahlung. We study the relative importance of these radiative processes across several accretion rates and black hole spins, and discuss the SEDs computed for \mes{} and \sgra{}.

The paper is structured as follows: in \S\ref{sec:techniques}, we describe the relevant equations and numerical methods, particularly the implementation of bremsstrahlung in \code{grmonty}. In \S\ref{sec:results}, we discuss the importance of bremsstrahlung for various accretion rates for the axisymmetric \radgrmhd{} simulations. We then present the computed SEDs for \mes{}{} and \sgra{}, and compare them to observations. We conclude with final remarks and possibilities for future work in \S\ref{sec:conclusion}. \replaced{A brief review and comparison of bremsstrahlung formul\ae{} in the literature can be found in the appendix.}{In the appendix, we briefly review and compare bremsstrahlung formul\ae{} found in the literature.}

\section{Techniques}\label{sec:techniques}
The radiative transfer calculation is done by post-processing the fluid calculation. Both calculations assume that the plasma is composed of ionized hydrogen, is charge-neutral, and has a thermal electron velocity distribution. Here, we define the emission coefficient $j_\nu$ as the power emitted per unit volume per unit frequency per unit solid angle, and the cooling rate $\Lambda$ as the power emitted per unit volume.

The fluid data were produced using two different codes: \code{ebhlight} and \code{harm}. \code{ebhlight} evolves the \radgrmhd{} equations with frequency-dependent radiative transfer, including the effects of synchrotron radiation and inverse Compton scattering. \code{ebhlight} is an extension of \code{bhlight} \citep{Ryan2015} that \deleted{also }tracks electron and ion temperatures independently according to the electron thermodynamics model of \cite{Ressler2015}, and uses the \cite{Howes2010} turbulent cascade model for electron heating. Both \code{bhlight} and \code{ebhlight} are based on the GRMHD code \code{harm} \citep{Gammie2003} and the radiative transfer code \code{grmonty} \citep{Dolence2009}.

\replaced{Three sets of fluid simulations of standard and normal evolution \citep{Narayan2012} accretion disks are used}{We use three sets of fluid simulations of standard and normal evolution \citep[SANE,][]{Narayan2012} accretion disks}. The first set \citep[][labeled \rmodellabel{}]{Ryan2017} contains four axisymmetric \radgrmhd{} \code{ebhlight} simulations. All of them have $m=10^{8}$ and $\spin=0.5$, but their time-averaged accretion rates $\equiv \av{\mdot}{t}$ are different and range from $\av{\mdot}{t}\approx10^{-8}$ to $\av{\mdot}{t}\approx10^{-5}$.
These accretion rates cover the regime where radiative processes become relevant to the flow dynamics.

\begin{deluxetable}{ccccccc}
	\tablecaption{Summary of the three sets of simulations used in this work.\label{tab:models}}
	\tablehead{
		\colhead{Set label} & \colhead{$m$} & \colhead{$\av{\mdot}{t}$} & \colhead{$a_{*}$} & \colhead{\tpte{}}
	}
	\startdata
	\multirow{4}{*}{\rmodellabel{}} & \multirow{4}{*}{$10^{8}$} & $1.1\times10^{-8}$ & \multirow{4}{*}{$0.5$} & \multirow{4}{*}{N/A}\\
	 & & $1.2\times10^{-7}$ & & \\
	 & & $9.3\times10^{-7}$ & & \\
	 & & $1.0\times10^{-5}$ & & \\\hline
	\multirow{4}{*}{\mes{}} & \multirow{2}{*}{$3.3\times10^{9}$} & $2.2\times10^{-5}$ & $0.5$ & \multirow{4}{*}{N/A}\\
	& & $8.2\times10^{-6}$ & $0.9375$ & \\
	& \multirow{2}{*}{$6.2\times10^{9}$} & $9.2\times10^{-6}$ & $0.5$ & \\
	& & $5.2\times10^{-6}$ & $0.9375$ & \\\hline
	\multirow{3}{*}{\sgra{}} & \multirow{3}{*}{$4.05\times10^{6}$} & $1.4\times10^{-8}$ & \multirow{3}{*}{$0.9375$} & 1\\
	& & $4.0\times10^{-8}$ & & 3\\
	& & $6.4\times10^{-7}$ & & 10\\
	\enddata
	\tablecomments{The first two sets are axisymmetric \radgrmhd{} \texttt{ebhlight} simulations. The third is a 3D GRMHD \texttt{harm} simulation scaled for \sgra{}. For more information about the first set (labeled R), see \cite{Ryan2017}, and for the second set (labeled \mes{}), see \cite{Ryan2018}.}
\end{deluxetable}

The second set \citep[][labeled \mes{}]{Ryan2018} contains four axisymmetric \radgrmhd{} \code{ebhlight} simulations of \mes{}, corresponding to the four possible combinations of two masses ($m=3.3\times10^{9}$ and $m=6.2\times10^{9}$) and two spins ($\spin=0.5$ and $\spin=0.9375$). For each simulation, the accretion rate is \replaced{chosen as required to match the observed flux density at $\SI{230}{\giga\hertz}$ \citep{Doeleman2012}}{such that the flux density at \SI{230}{\giga\hertz} matches the observed value \citep{Doeleman2012}}.

Both sets of \code{ebhlight} simulations used axisymmetrized 3D GRMHD simulation data as initial conditions. This procedure alleviates some limitations of axisymmetry, such as the long integration times required to achieve viscous electron heating equilibrium at larger radii. The SEDs obtained from these simulations are then time-averaged between $t=600GM/c^{3}$ and $t=1000GM/c^{3}$ with a $5GM/c^3$ cadence.

The third set (labeled \sgra{}) contains a 3D GRMHD \code{harm} simulation scaled for \sgra{} ($m=4.05\times10^{6}$, $\spin=0.9375$). \replaced{A single fluid temperature is assumed for the duration of the \code{harm} GRMHD simulation, after which a fixed temperature ratio between protons and electrons ($\equiv\ms{T}{p}/\ms{T}{e}$) is assigned in the radiative transfer calculation}{The GRMHD simulation assumes a single fluid temperature, and then the radiative transfer calculation assigns a fixed temperature ratio between protons and electrons ($\equiv\ms{T}{p}/\ms{T}{e}$)}. \replaced{This work uses different}{Here we consider} three values for this ratio (1, 3, and 10). \replaced{The SEDs were computed}{We compute SEDs} at an observing angle of $\pi/3$ and then time-average between $t=2000GM/c^{3}$ and $t=10^4 GM/c^{3}$ with a $100GM/c^3$ cadence. For every value of $\ms{T}{p}/\ms{T}{e}$, we normalize the GRMHD model so that the time-averaged \SI{230}{\giga\hertz} flux in our SEDs matches the observed $\SI{2.4}{\jansky}$ \citep{Doeleman2008} to within $\sim1\%$.

Table \ref{tab:models} shows a summary of all simulations.

The fluid data from these simulations are post-processed using \code{grmonty} \citep{Dolence2009}, a relativistic Monte Carlo radiative transfer code. The code computes an SED from a single time slice of the fluid data using the ``fast light'' approximation, \replaced{in which photons propagate}{i.e. propagating photons} through the computational domain while the fluid variables are held constant in time. The code accounts for synchrotron radiation, inverse Compton scattering, and bremsstrahlung (see \S\ref{sec:techniques:bremss}). In addition to the total SED, \code{grmonty} records the SED produced by each radiative process individually. Photons produced by synchrotron or bremsstrahlung that get scattered as they propagate through the plasma are marked as inverse Compton scattering photons.

\subsection{Bremsstrahlung}\label{sec:techniques:bremss}
We consider bremsstrahlung emission from electron-ion and electron-electron encounters in an ionized hydrogen plasma. Fitting formul\ae{} for the bremsstrahlung cooling rate \citep{Svensson1982} show that in the high temperature limit, electron-electron bremsstrahlung contributes two thirds of the total emitted power, i.e., $\coolingrate{br,ee}/\coolingrate{br,ei}=3\F{ee}/\lp8\pi\F{ei}\rp\sim2$. Here $\thte\equiv\kb\ms{T}{e}/\ms{m}{e}c^{2}$ is the dimensionless electron temperature, where $\ms{T}{e}\equiv$ electron temperature, $\ms{k}{B}\equiv$ Boltzmann constant, $\ms{m}{e}\equiv$ electron mass, and $c\equiv$ speed of light. The functions \F{ee} and \F{ei} are \deleted{the }given in \cite{Stepney1983} and reproduced in the appendix. From these formul\ae{}, the temperature at which both types of bremsstrahlung have the same cooling rate is $\thte\sim0.67$. The space-averaged $\thte$ in the simulations used here is of order unity, so electron-electron bremsstrahlung should be taken into account. The emission from outside the simulation domain ($r>200GM/c^{2}$) is neglected.

Many bremsstrahlung emission coefficient formul\ae{} exist in the literature. The appendix summarizes and compares some of these. For electron-ion bremsstrahlung, \code{grmonty} uses the emission coefficient \citep[see, e.g.,][]{RL1979}
\begin{equation}
\jnu{br,ei}=\frac{8 \ms{q}{e}^{6}}{3\ms{m}{e}^{2}c^{4}}\sqrt{\frac{2\pi}{3}}\thte^{-1/2}\ms{n}{e}n_{i}e^{-h\nu/\ms{k}{B}\ms{T}{e}}\gaunt{ei},
\end{equation}
where $\ms{q}{e}\equiv$ elementary charge, $\ms{n}{e}\equiv$ electron number density, $\ms{n}{i}\equiv$ ion number density, $h\equiv$ Planck constant, and $\gaunt{ei}$ is the thermally-averaged electron-ion Gaunt factor given as tabulated values in \citet{VanHoof2015}. \added{This Gaunt factor combines an exact non-relativistic calculation \citep{VanHoof2014} with a relativistic calculation in the Born approximation, thus spanning a large parameter space accurately.} \code{grmonty} interpolates these values using $\ms{T}{e}$ and $\nu$ as independent variables. For electron-electron bremsstrahlung, \code{grmonty} uses the emission coefficient of \citet{Nozawa2009},
\begin{equation}
\jnu{br,ee}=\frac{4\pi\ms{q}{e}^{6}}{3\ms{m}{e}^{2}c^{4}}\thte^{1/2}\ms{n}{e}^{2}e^{-h\nu/\ms{k}{B}\ms{T}{e}}\gaunt{ee},
\end{equation}
where $\gaunt{ee}$ is the thermally-averaged electron-electron Gaunt factor given in \citet{Nozawa2009} as a piecewise fitting formula. \added{This formula combines a non-relativistic calculation \citep{Itoh2002} with calculations where relativistic effects and/or Coulomb corrections are important.} When \code{grmonty} encounters points in parameter space that are outside the domains of these two numerical calculations, it uses the nearest point inside the domains. These points are rare, and the spectrum is mostly insensitive to the way they are handled. We use Kirchhoff's law to account for bremsstrahlung absorption.

\section{Results}\label{sec:results}

\begin{figure}[b]
\includegraphics[width=\columnwidth]{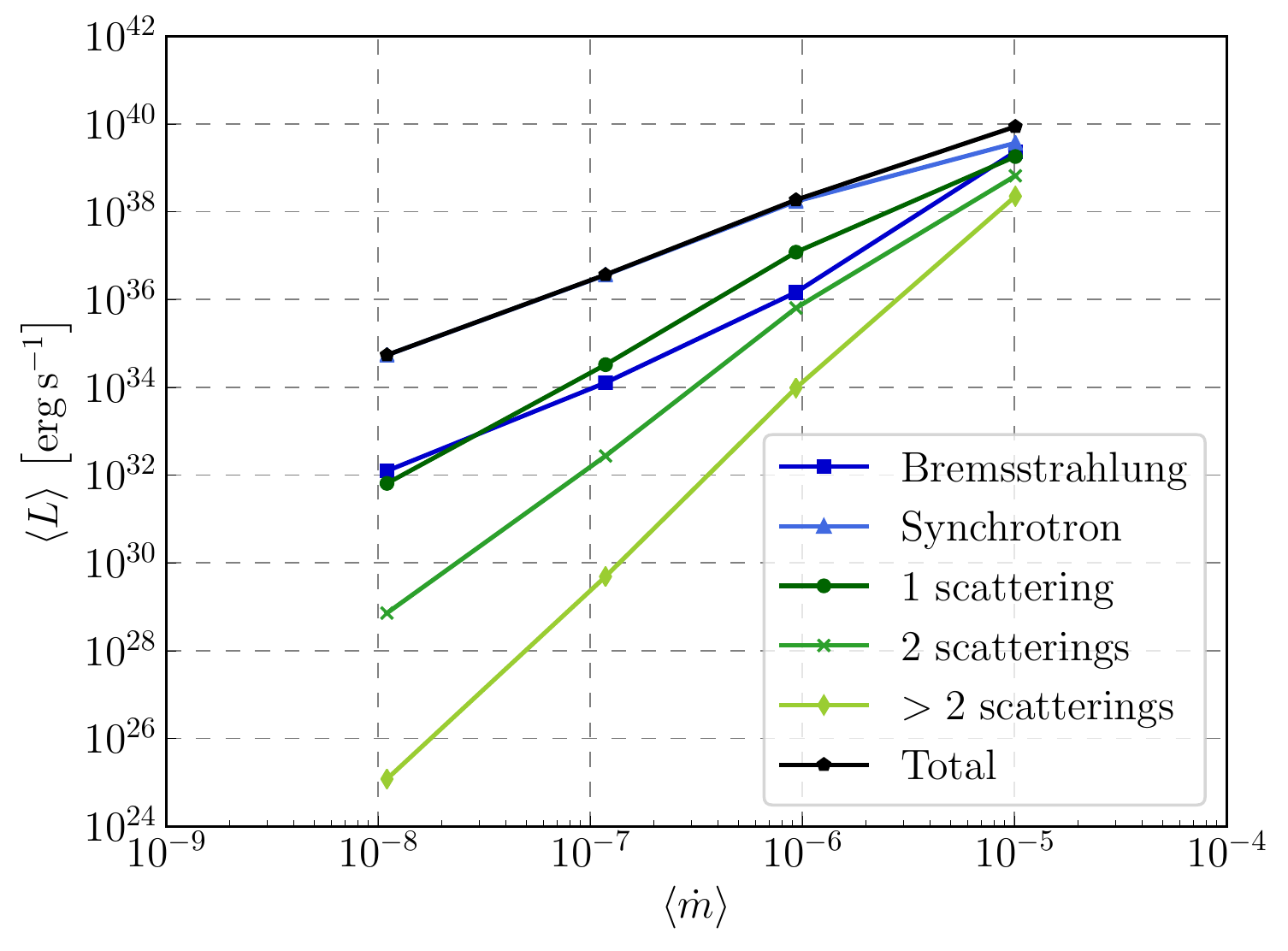}
\caption{\added{Time-averaged total} luminosity for each radiative process in model set \rmodellabel{} ($m=10^{8}$, $\spin=0.5$) as a function of accretion rate. While synchrotron radiation dominates at lower accretion rates, bremsstrahlung and inverse Compton scattering become increasingly important as the accretion rate increases.}
\label{fig:PowerR2017}
\end{figure}

\begin{figure*}[t!]
\gridline{\fig{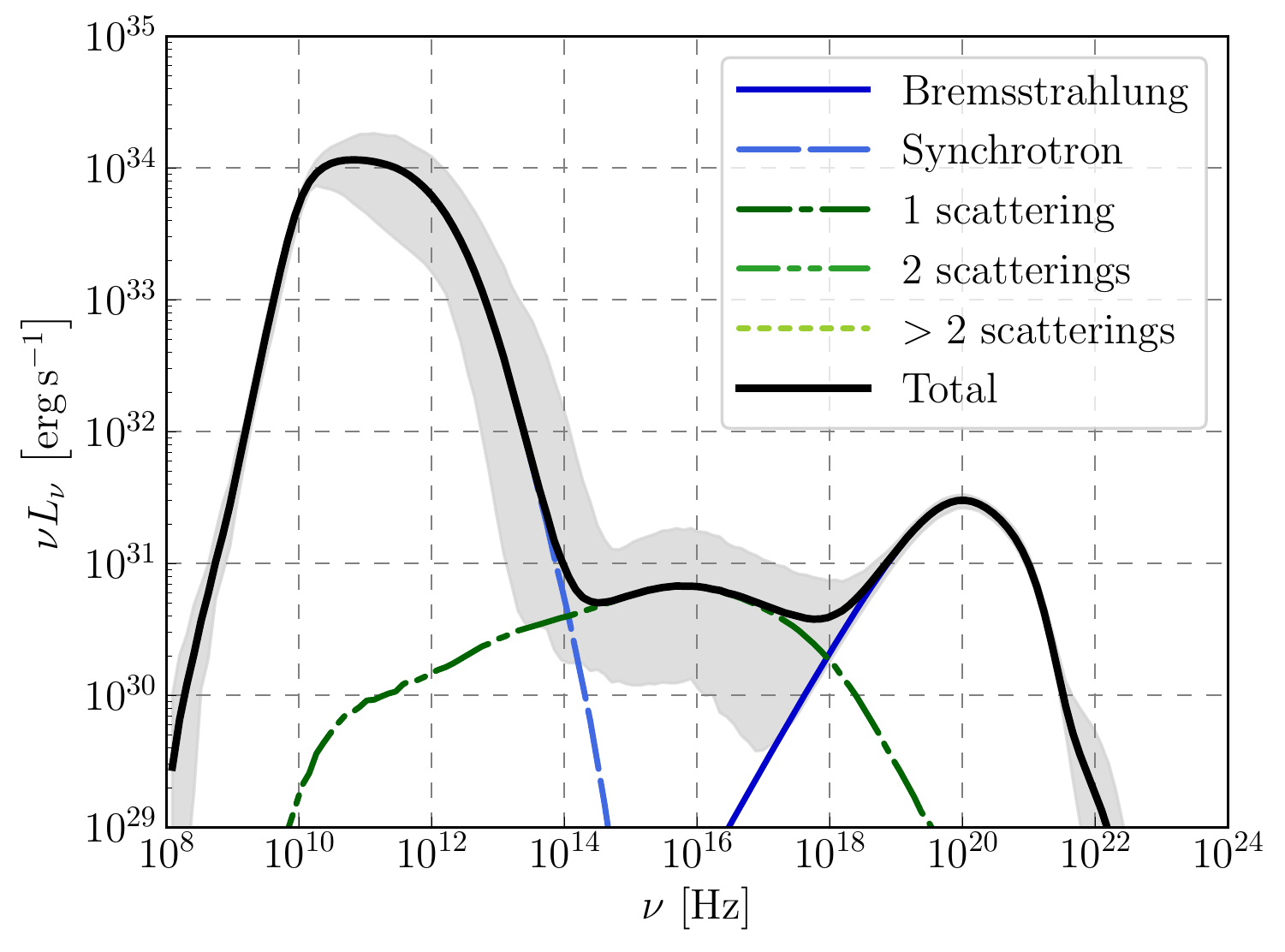}{\spectrumgridsize{}}{$\av{\mdot}{t}\approx10^{-8}$}
     \fig{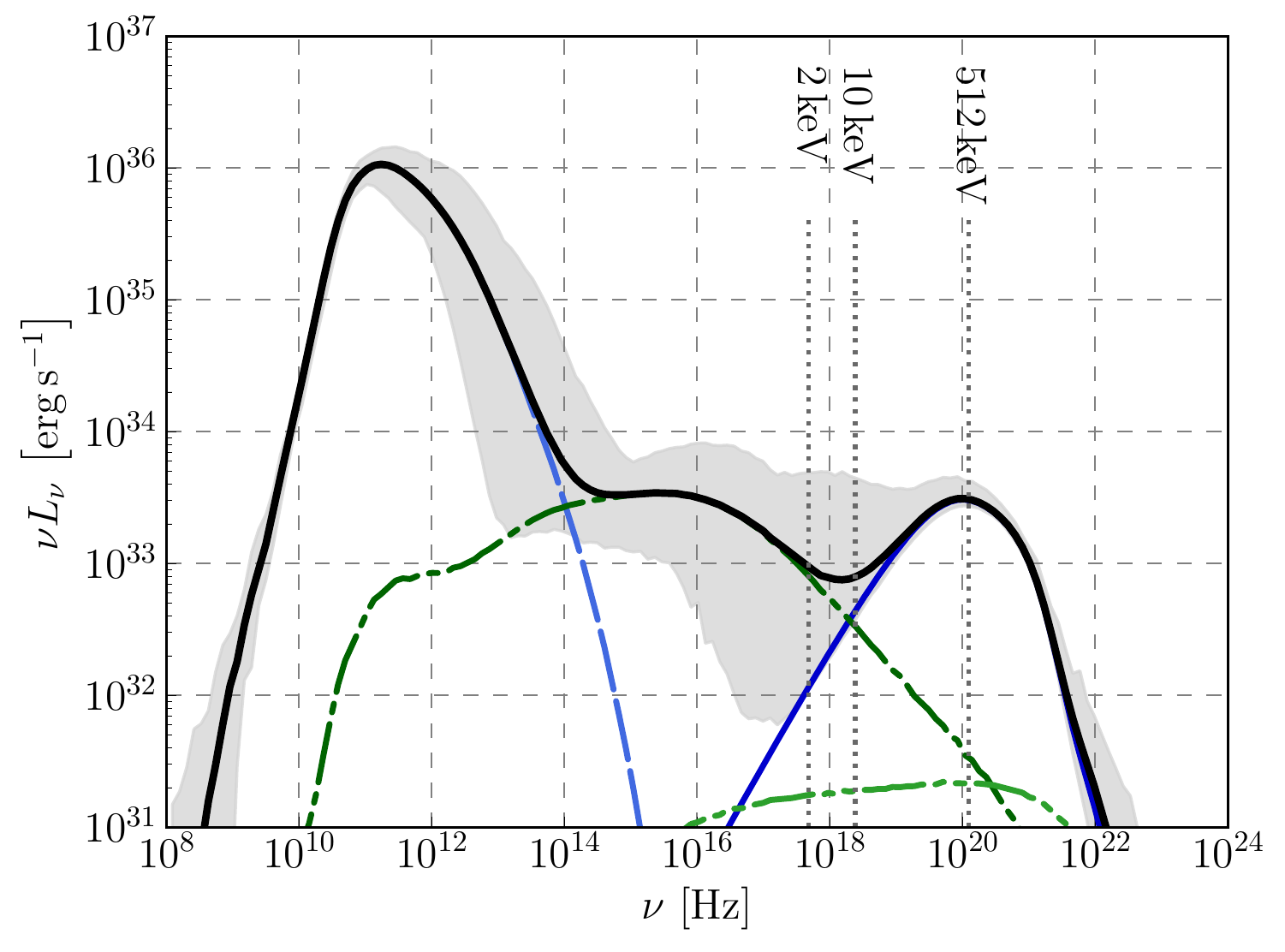}{\spectrumgridsize{}}{$\av{\mdot}{t}\approx10^{-7}$}}
\gridline{\fig{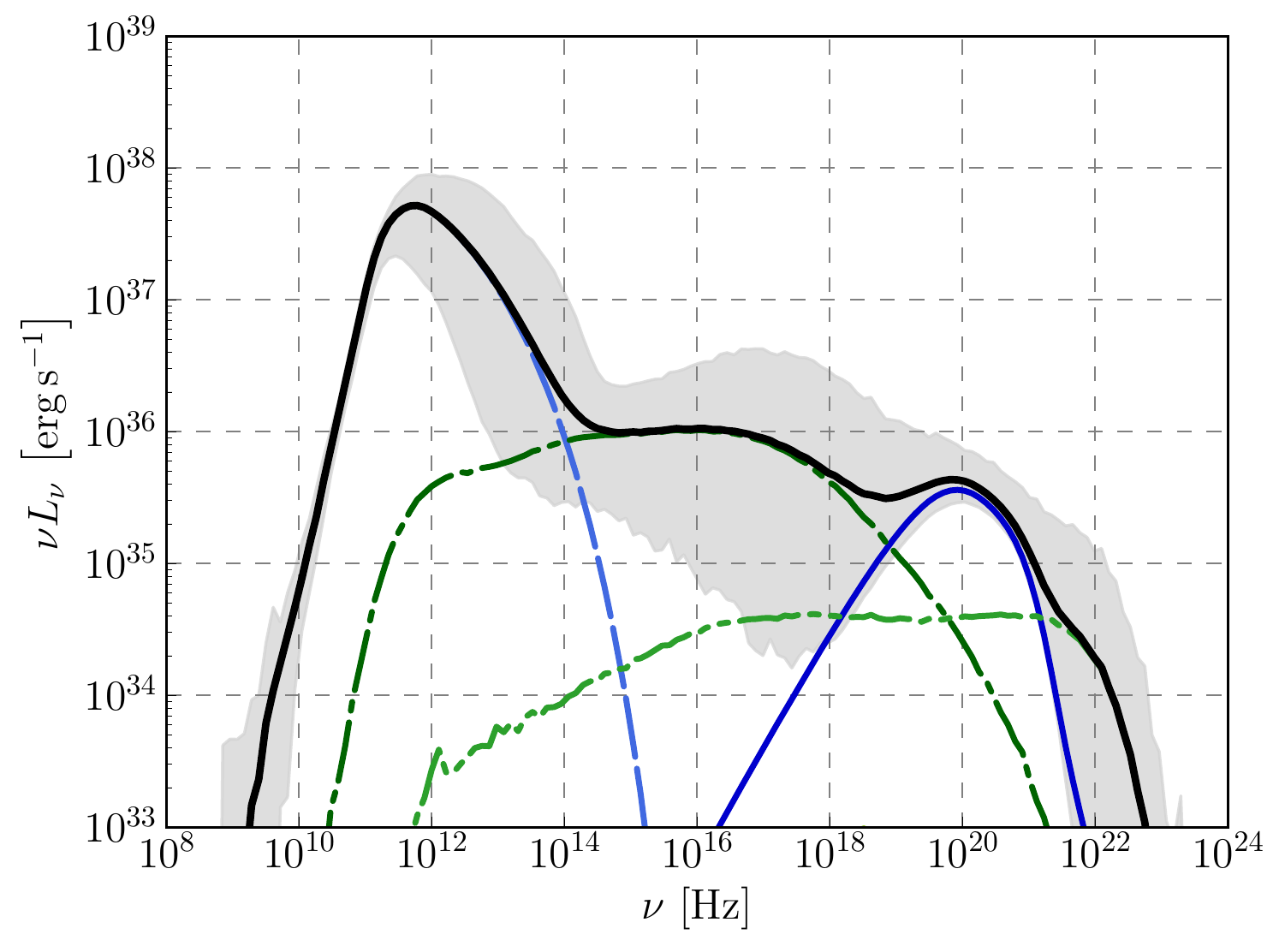}{\spectrumgridsize{}}{$\av{\mdot}{t}\approx10^{-6}$}
	\fig{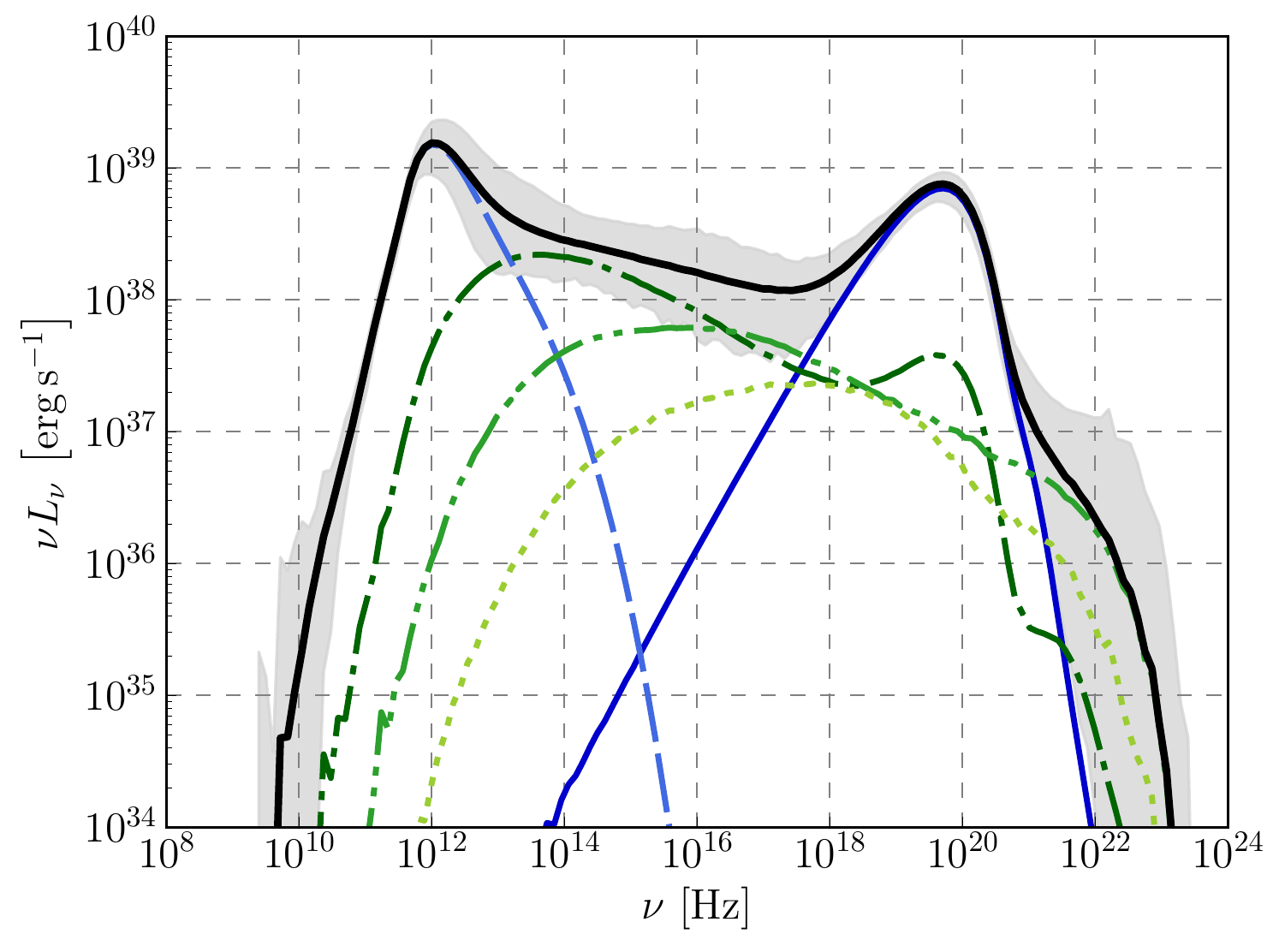}{\spectrumgridsize{}}{$\av{\mdot}{t}\approx10^{-5}$}}
\caption{Time-averaged spectral energy distributions (SEDs) for the set \rmodellabel{} models ($m=10^{8}$, $\spin{}=0.5$). Each panel corresponds to one accretion rate. The shaded region around the total SED shows the variability throughout the time-averaging period. Bremsstrahlung dominates the SED near $\nu\sim\SI{e20}{\hertz}$ ($h\nu\sim\SI{512}{\kilo\electronvolt}$) at all accretion rates.}
\label{fig:SEDs:R2017}
\end{figure*}

We report quantities that have been time-averaged (the time average of $f$ is denoted by $\av{f}{t}$) over the interval when the disk is in statistical steady state. In \S\ref{sec:results:R2017}, we discuss general results about bremsstrahlung as a function of accretion rate. In \S\ref{sec:results:M87SgrA}, we show computed SEDs for \mes{} and \sgra{} and compare them to observations.

\subsection{Bremsstrahlung as a function of \mdot}\label{sec:results:R2017}

Figure \ref{fig:PowerR2017} shows the \added{time-averaged total} luminosity produced by each radiative process as a function of accretion rate for the models in set \rmodellabel{}. Synchrotron produces most of the luminosity at low accretion rates by a few orders of magnitude, but other radiative processes become increasingly important as the accretion rate increases. At $\av{\mdot}{t}\approx10^{-5}$, synchrotron, bremsstrahlung, and inverse Compton scattering produce about $40\%$, $30\%$, and $30\%$ of the total luminosity, respectively.

\begin{figure*}
\gridline{\fig{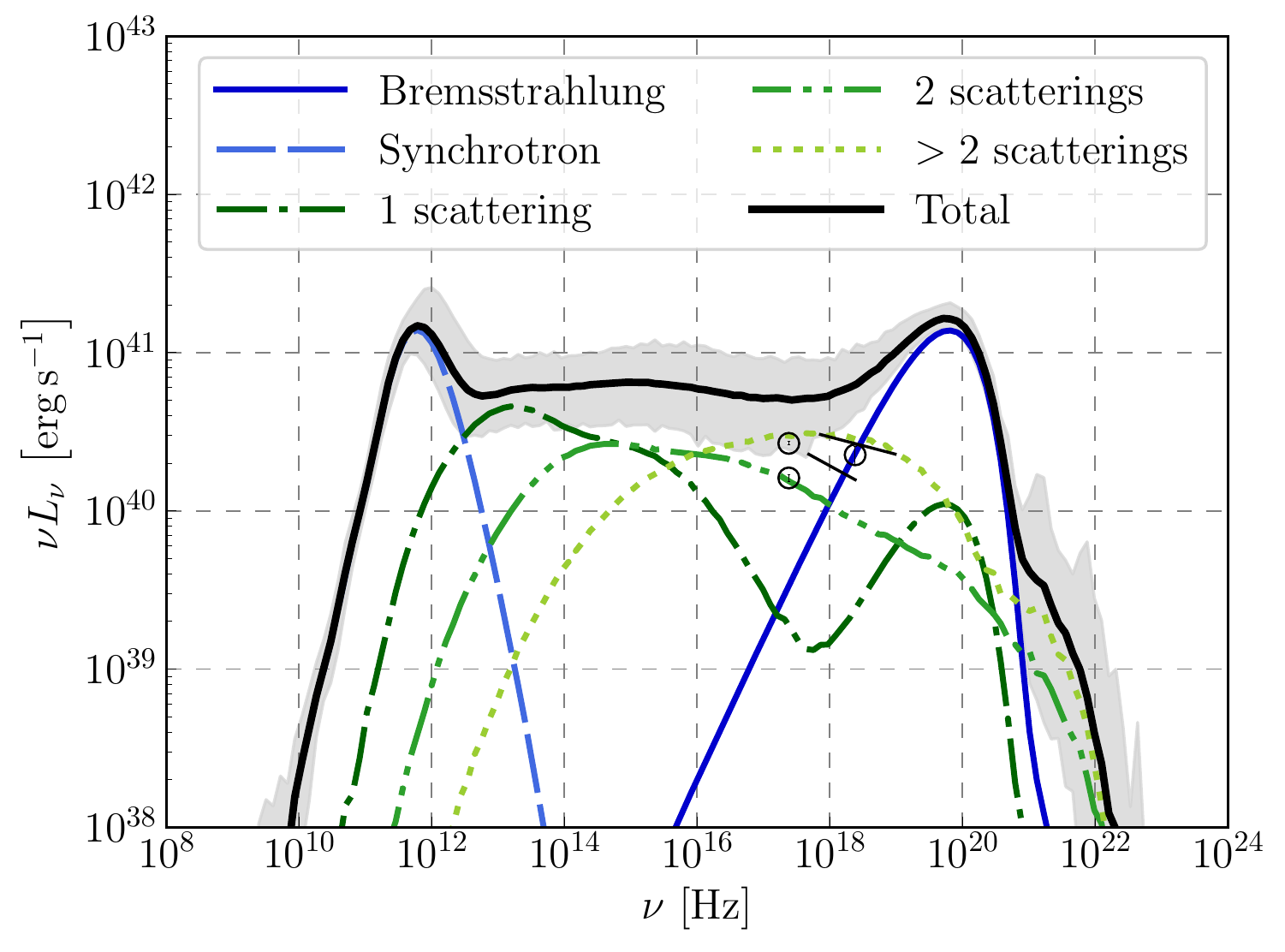}{\spectrumgridsize}{$m=3\times10^{9}$, $a_{*}=0.5$, $\av{\mdot}{t}=2.2\times10^{-5}$}
     \fig{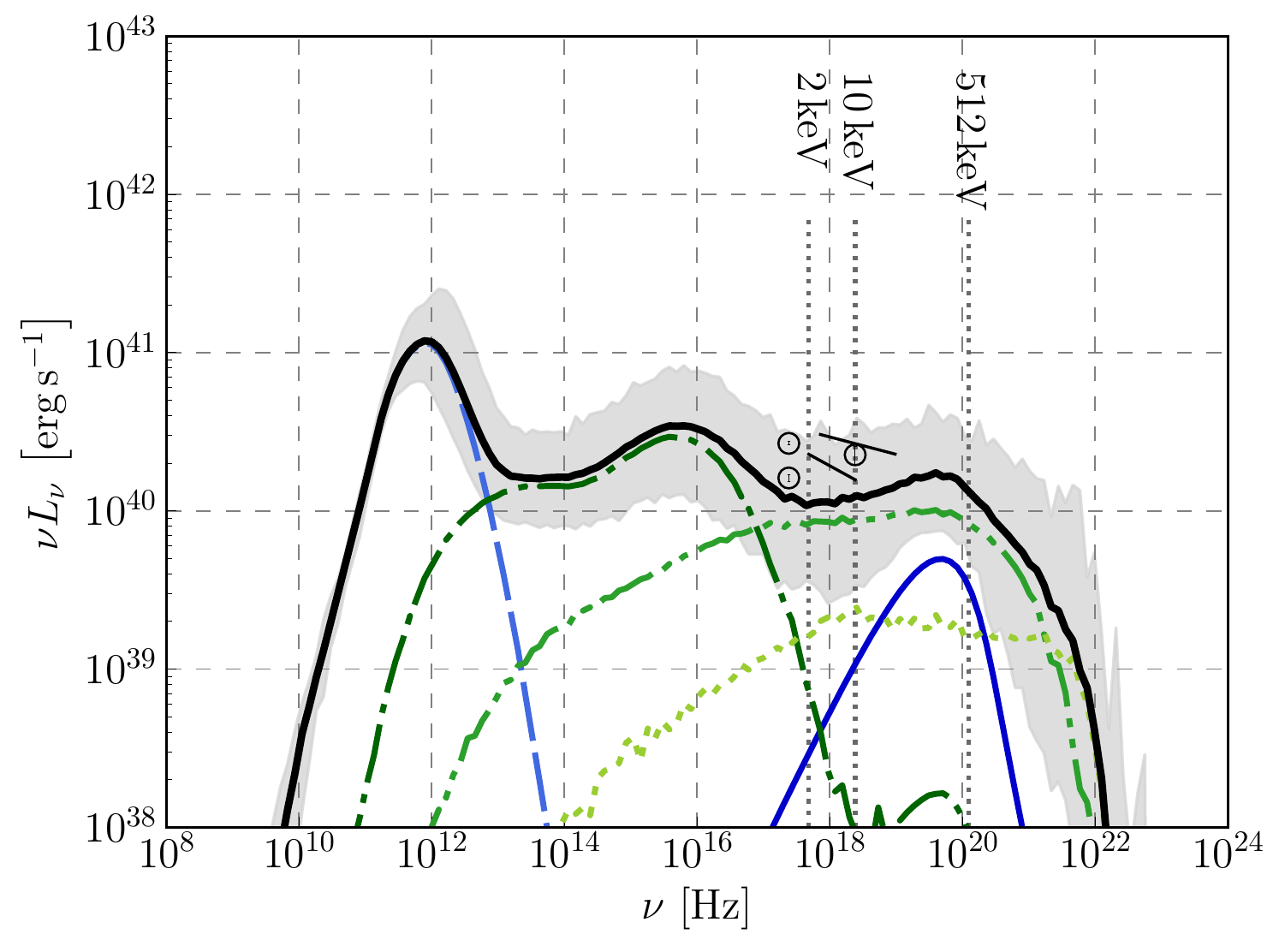}{\spectrumgridsize}{$m=3\times10^{9}$, $a_{*}=0.9$, $\av{\mdot}{t}=8.2\times10^{-6}$}
     }
\gridline{\fig{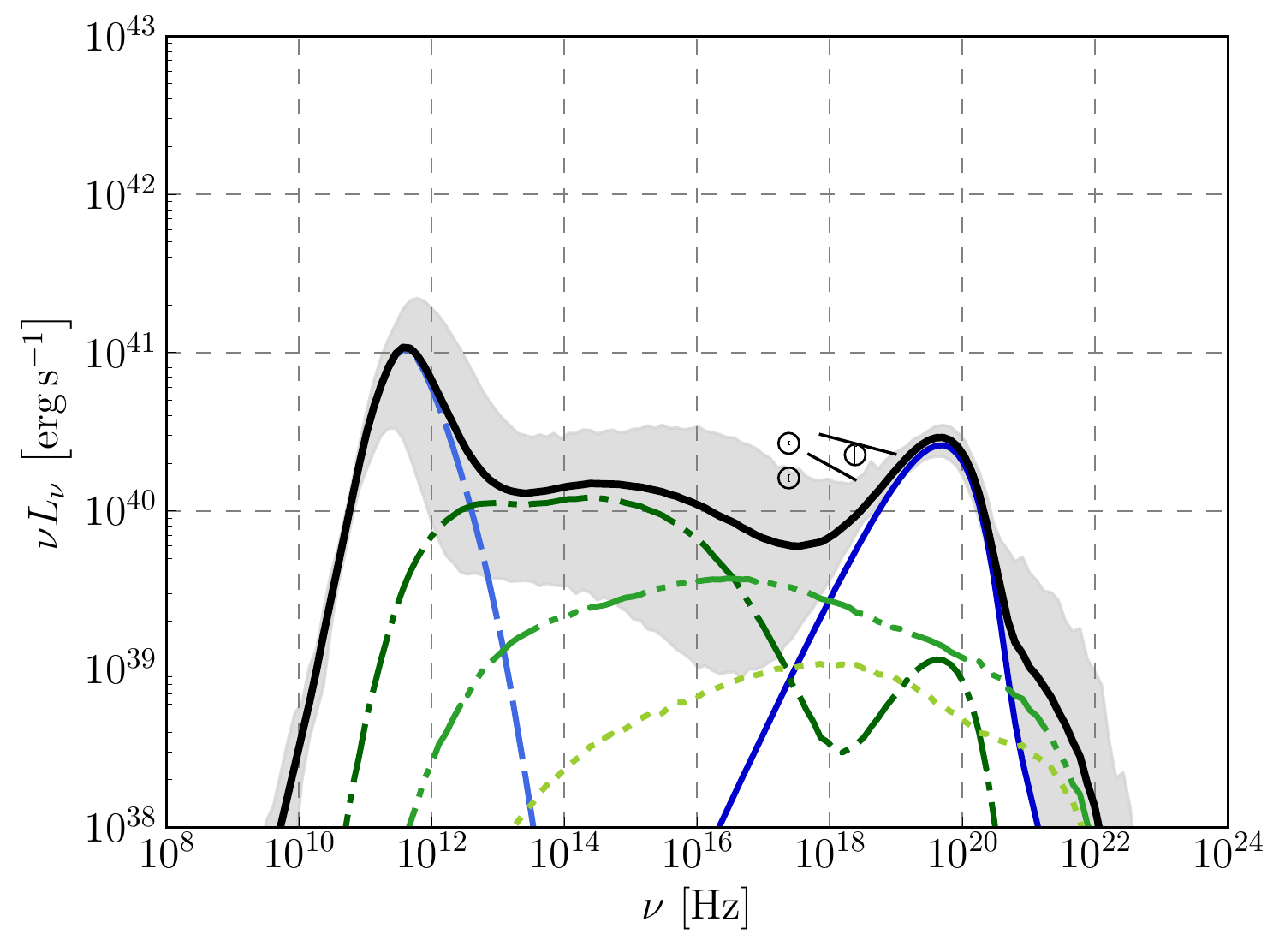}{\spectrumgridsize}{$m=6\times10^{9}$, $a_{*}=0.5$, $\av{\mdot}{t}=9.2\times10^{-6}$}
	\fig{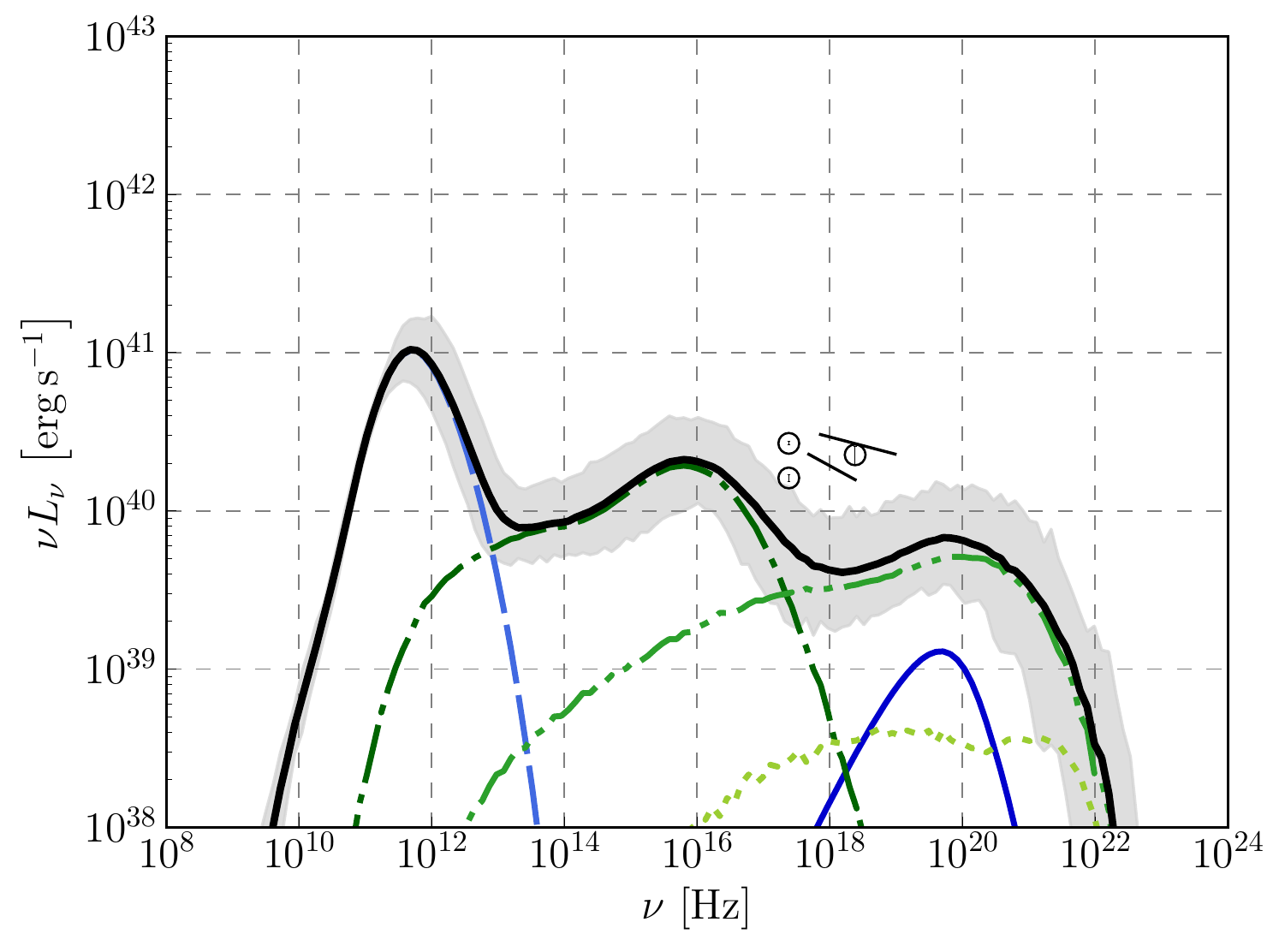}{\spectrumgridsize}{$m=6\times10^{9}$, $a_{*}=0.9$, $\av{\mdot}{t}=5.2\times10^{-6}$}
}
\caption{Time-averaged computed spectral energy distributions (SEDs) for the \mes{} models. The shaded region around the total SED shows the variability throughout the time averaging period ($600GM/c^{3}$ - $1000GM/c^{3}$). \xray{} observations \citep{DiMatteo2003, Prieto2016} are shown as circles, and their uncertainties as error bars inside the circles. Observed SEDs in the 2-\SI{10}{\kilo\electronvolt} \citep{DiMatteo2003} and 3-\SI{40}{\kilo\electronvolt} \citep{Wong2017} ranges are shown as solid black lines. Conversion from $\si{\jansky}$ to \nuLnu assumed isotropic emission and a distance $D=\SI{16.7}{\mega\parsec}$ to \mes{}. The model with $m=3\times10^{9}$ and $\spin=0.5$ overproduces \xrays{}. Bremsstrahlung dominates near $\nu\sim\SI{e20}{\hertz}$ when $\spin=0.5$, and inverse Compton scattering dominates when $\spin=0.9375$.\label{fig:SEDs:M87}}
\end{figure*}

\replaced{A slight change in the scaling of the bremsstrahlung luminosity with \mdot{} can be seen in Figure \ref{fig:PowerR2017}. These change occurs at $\av{\mdot}{t}\approx10^{-6}$, the point at which radiation effects on flow dynamics become significant \citep{Ryan2017}. For $\av{\mdot}{t}\leq10^{-6}$, the bremsstrahlung luminosity agrees with the ADAF scaling, i.e. $\av{\Lum{br}}{t}\propto\av{\mdot}{t}^{2}$. However, when $\av{\mdot}{t}\geq10^{-6}$, the bremsstrahlung luminosity scaling changes to approximately $\av{\Lum{br}}{t}\propto\av{\mdot}{t}^{3}$. This change in scaling is at least partially caused by increasing Coulomb heating of electrons in the outer part of the disk as the accretion rate increases. In our models, as a consequence of the limited extent of the disk, the bremsstrahlung cooling rate peaks at $r\sim40GM/c^2$. While thermal equilibration is a concern here, we find that $\sigma_{\thte}/\av{\thte}{t}\lesssim0.1$ near the peak in the bremsstrahlung cooling rate.}{The ratio of bremsstrahlung and synchrotron luminosities scales roughly as $\coolingrate{br}/\coolingrate{synch}\propto\ms{n}{e}^2 \thte/\lp\ms{n}{e}B^2\thte^2\rp=\ms{n}{e}/\lp B^2\thte\rp$, which is constant in the low-accretion rate limit (because $\ms{n}{e}\propto\mdot$, $B\propto\mdot^{1/2}$, $\thte$ constant). Between our $\av{\mdot}{t}\approx10^{-8}$ and $\av{\mdot}{t}\approx10^{-7}$ simulations, the ratio between the bremsstrahlung and synchrotron luminosities increases by a factor of $1.5$. In contrast, between $\av{\mdot}{t}\approx10^{-6}$ and $\av{\mdot}{t}\approx10^{-5}$, this ratio increases by a factor of 75, which shows that bremsstrahlung is becoming more important as the accretion rate increases. The ratio changes partly because the synchrotron luminosity scales sub-quadratically with \av{\mdot{}}{t}. Most synchrotron radiation is produced in the inner part of the disk, and in our models, electrons in this region are colder at higher \av{\mdot}{t} \citep{Ryan2017}. Additionally, the scaling of the bremsstrahlung luminosity with \av{\mdot{}}{t} changes slightly at $\av{\mdot}{t}\approx10^{-6}$ as the effect of radiation on flow dynamics becomes significant. For $\av{\mdot}{t}\leq10^{-6}$, the bremsstrahlung luminosity agrees with the low-\mdot{} scaling, i.e. $\av{\Lum{br}}{t}\propto\av{\mdot}{t}^{2}$. However, when $\av{\mdot}{t}\geq10^{-6}$, the bremsstrahlung luminosity scaling changes to approximately $\av{\Lum{br}}{t}\propto\av{\mdot}{t}^{3}$. This change in scaling is at least partially caused by increasing Coulomb heating of electrons at $r\gtrsim15GM/c^2$ as the accretion rate increases \citep{Ryan2017}. In our models, as a consequence of the limited extent of the disk, the bremsstrahlung cooling rate peaks at $r\sim40GM/c^2$. While thermal equilibrium is a concern here, we find that near the spatial peak of the bremsstrahlung cooling rate $\sigma_{\thte}/\av{\thte}{t}\lesssim0.1$, where $\sigma$ represents the standard deviation in time. Finally, the bremsstrahlung absorption optical depth near the peak frequency $\sim\SI{e20}{\hertz}$ is very small, so photons near the peak frequency can escape almost freely at all accretion rates.}

Figure \ref{fig:SEDs:R2017} shows the computed SEDs as a function of accretion rate for models in set \rmodellabel{}. Bremsstrahlung dominates the SED around $\nu\sim\SI{e20}{\hertz}$ at all accretion rates. However, the relative importance of bremsstrahlung and inverse Compton scattering is sensitive to at least black hole spin (see \S\ref{sec:results:M87SgrA}). The computed bremsstrahlung luminosity near the peak (between $\SI{e18}{\hertz}$ and $\SI{e21}{\hertz}$) satisfies $\sigma_{L}/\av{L}{t}\sim0.1$ for all accretion rates. On the other hand, inverse Compton scattering has $\sigma_{L}/\av{L}{t}\sim1$ at all accretion rates except at $\av{\mdot}{t}\approx10^{-5}$, where $\sigma_{L}/\av{L}{t}\sim0.5$.

\subsection{\texorpdfstring{\mes{}}{M87*} and \texorpdfstring{\sgra{}}{SgrA*}}\label{sec:results:M87SgrA}
Figure \ref{fig:SEDs:M87} shows the computed SEDs for \mes{}\replaced{ plotted against}{, as well as} \xray{} observations \citep{DiMatteo2003, Prieto2016,Wong2017}. These observations measured flux density with an aperture radius of $\SI{0.4}{\arcsecond}$, which means they exclude emission from the jet at larger scales. Assuming a distance $D=\SI{16.7}{\mega\parsec}$ to \mes{}, they captured the emission from a region of radius $\sim10^{5}GM/c^{2}$, which is considerably larger than \replaced{the outer radius of the simulations $200GM/c^{2}$ used in this work}{$200GM/c^{2}$, the outer radius of our simulations}. Although most of the emission comes from the inner region of the accretion flow, this discrepancy in simulation and observation sizes means the observations should be interpreted as upper limits.

At $\spin=0.5$, bremsstrahlung dominates the SED near $\nu\sim\SI{e20}{\hertz}$, \replaced{which is consistent}{in agreement} with the results from model \rmodellabel{}. However, when $\spin=0.9375$, the upscattered synchrotron photons dominate this part of the SED, and the bremsstrahlung peak is approximately one order of magnitude below. Accordingly, when $\spin=\lb0.5,0.9375\rb$, bremsstrahlung contributes $\sim\lb20\%,1\%\rb$ of the total luminosity. This difference between low and high spin models \replaced{arises from the different temperature and density profiles between them}{arises from their different temperature and density profiles}. For all \added{\mes{}} models, the bremsstrahlung variability $\sigma_{L}/\av{L}{t}\sim0.2$, while $\sigma_{L}/\av{L}{t}\sim0.7$ for second order inverse Compton scattering. Since inverse Compton scattering is much more variable and dominates when $\spin=0.9375$, these results suggest that variability near $\nu\sim\SI{e20}{\hertz}$ could be a diagnostic of black hole spin. However, a significant part of the accretion disk in these simulations might not be in radiative and/or viscous equilibrium, so 3D studies are needed to know if this result is valid. Furthermore, \replaced{it is uncertain how the relationship between inverse Compton scattering and bremsstrahlung depends on the magnetic field structure}{the relationship between inverse Compton scattering and bremsstrahlung might depend on the magnetic field structure}. \replaced{This work uses}{We use} simulations of SANE accretion disks, as opposed to magnetically arrested disks \citep[MADs,][]{Narayan2003}\replaced{, which}{; the latter} have also been used to study \mes{} \citep{Chael2019,EHT2019P5}.

Fits to \xray{} observations of \mes{} provide an upper limit for the 2--\SI{10}{\kilo\electronvolt} luminosity of $\SI{5e40}{\erg\per\second}$ \citep{Bohringer2001}, $\SI{3.1e40}{\erg\per\second}$ \citep{DiMatteo2003}, and $\SI{4.4e40}{\erg\per\second}$ \citep{EHT2019P5}. The average luminosity $\av{L_{X}}{t}$ of the model with $m=3\times10^{9}$ and $\spin=0.5$ is a factor of $\approx3$ larger than observations. Accounting for variability, $\av{L_{X}}{t}-2\sigma_{L_{X}}=\SI{3e40}{\erg\per\second}$ is only marginally consistent. The fact that the observations encompass a larger part of the accretion flow and potentially unresolved sources strongly suggests this model is inconsistent with observations, as found in previous work \citep{Ryan2018}. The other models are less than an order of magnitude below this limit.

\begin{figure}[hbtp!]
\gridline{\fig{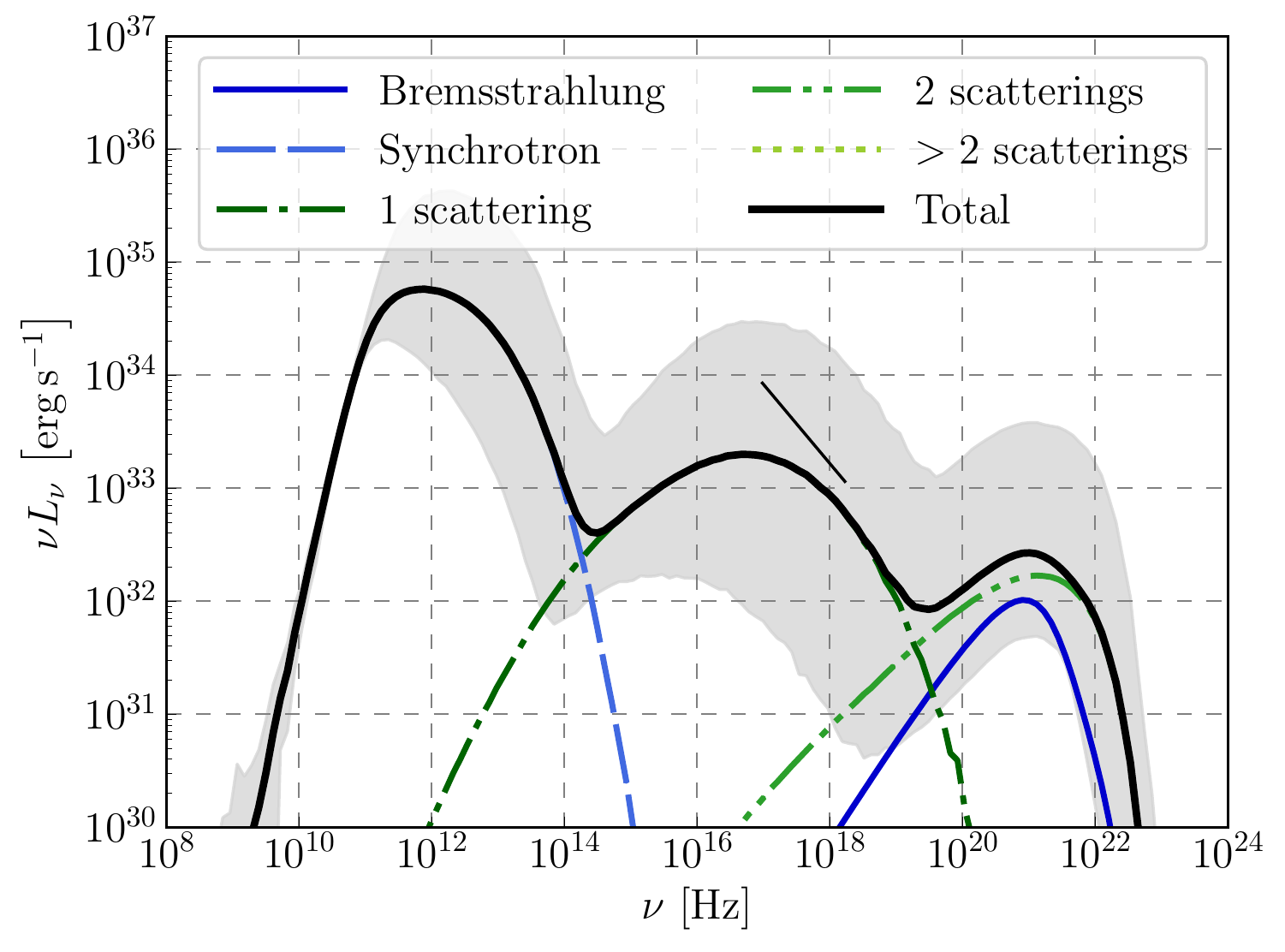}{0.92\columnwidth}{$\tpte=1$, $\av{\mdot}{t}=1.4\times10^{-8}$}}
\gridline{\fig{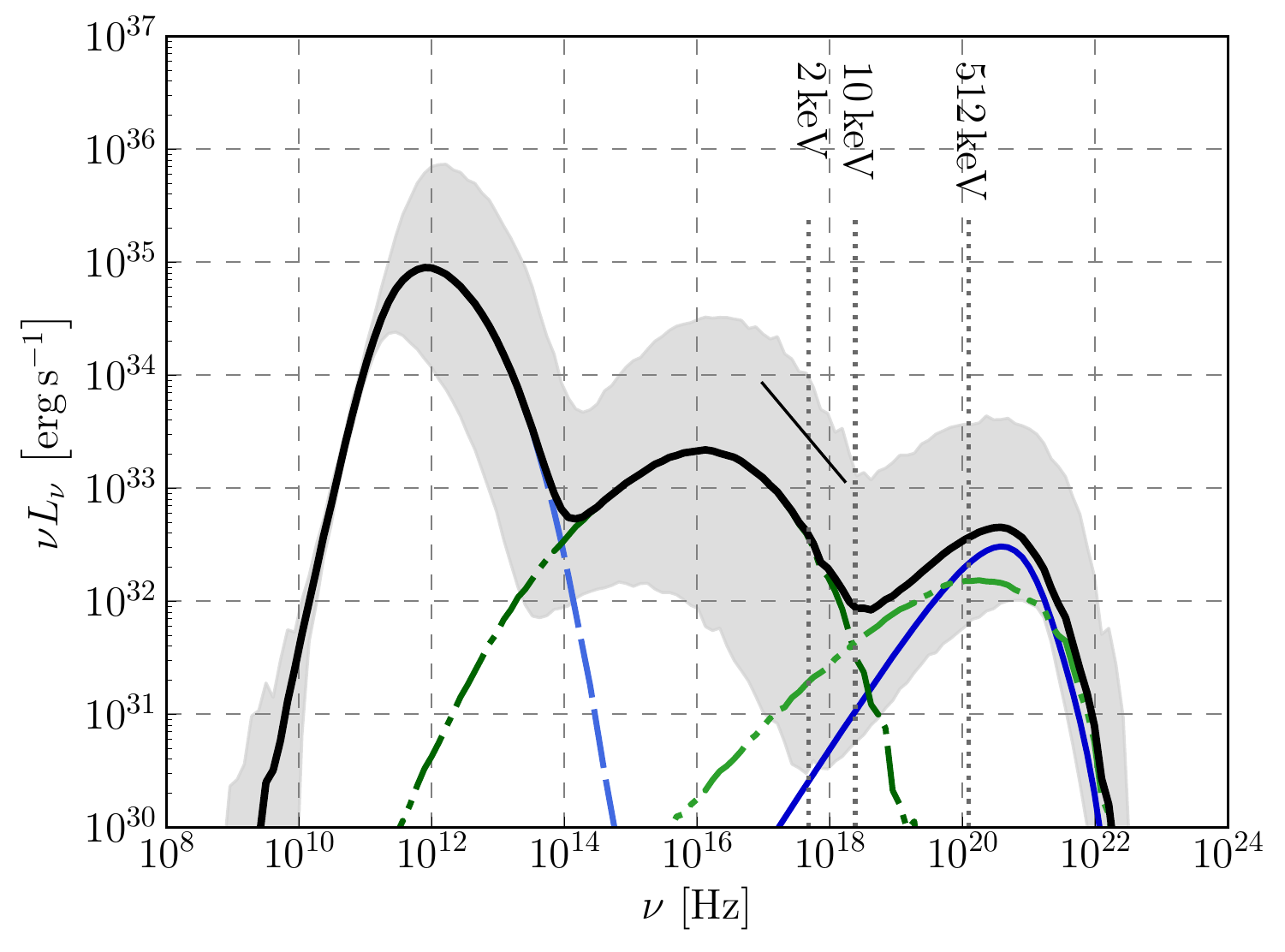}{0.92\columnwidth}{$\tpte=3$, $\av{\mdot}{t}=4\times10^{-8}$}}
\gridline{\fig{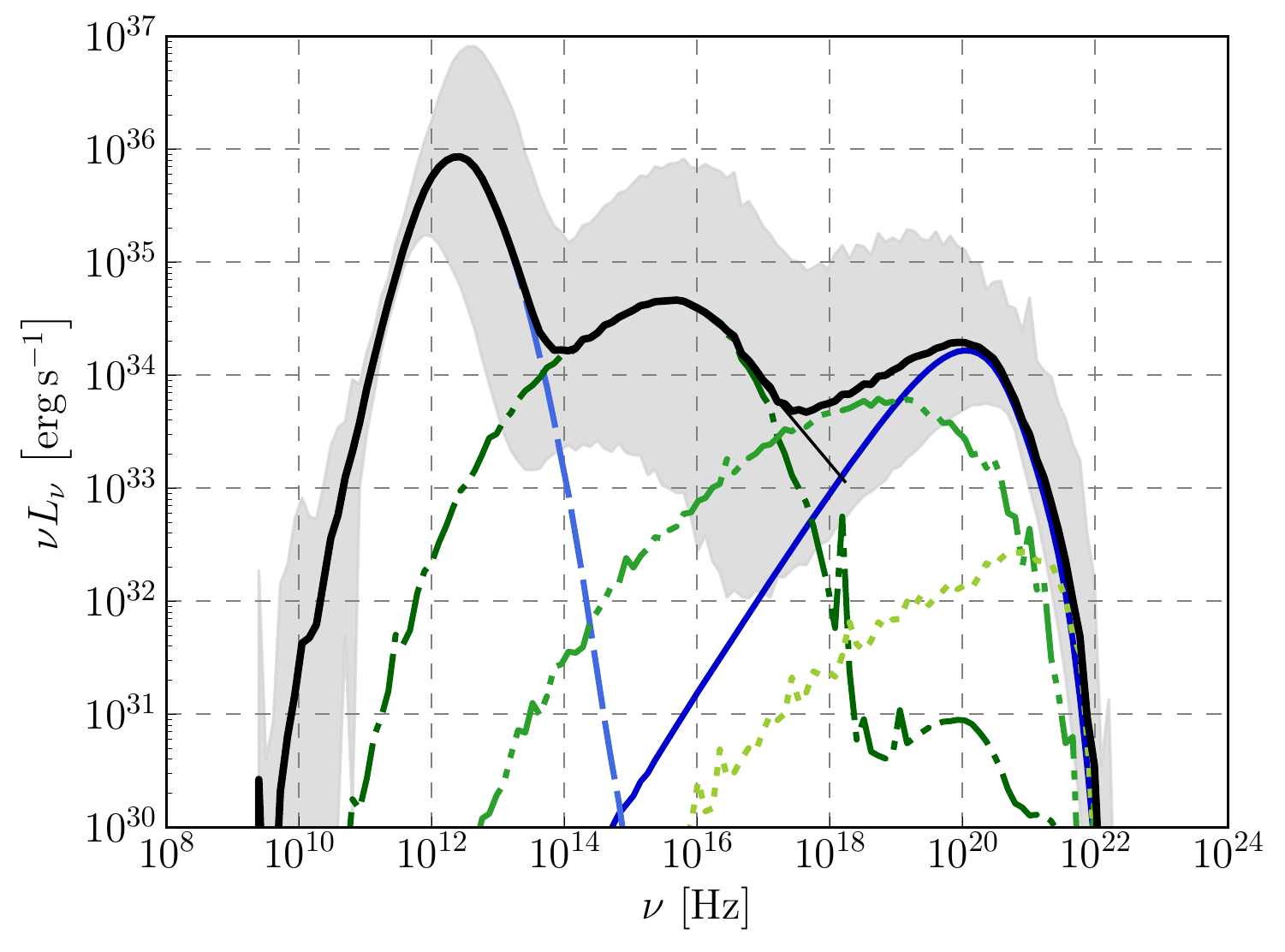}{0.92\columnwidth}{$\tpte=10$, $\av{\mdot}{t}=6.4\times10^{-7}$}}
\caption{Time-averaged computed spectral energy distributions (SEDs) for Sgr A* ($m=4.05\times10^{6}$, $\spin=0.9375$) at a viewing angle of $\pi/3$. The shaded region around the total SED shows the variability throughout the time averaging period. The SEDs are normalized to the observed \SI{230}{\giga\hertz} flux \citep{Doeleman2008}.}
\label{fig:SEDs:SgrA}
\end{figure}

In our GRMHD models of \sgra{} we find that at fixed $\tpte$ and accretion rate bremsstrahlung becomes increasingly important compared to inverse Compton scattering as the viewing angle increases. The same trend holds at fixed viewing angle with increasing \tpte{}. We hereafter focus on SEDs at a fixed viewing angle of $\sim\pi/3$ with respect to the black hole spin vector. These SEDs, shown in Figure \ref{fig:SEDs:SgrA}, were computed by recording $\nuLnu$ for photons in an angular bin between $\SI{47}{\degree}$ and $\SI{65}{\degree}$, corresponding to a solid angle $\Omega$, and then multiplying these $\nuLnu$ by $4\pi/\Omega$. The 2--\SI{10}{\kilo\electronvolt} luminosities are $\lb1.8,0.43,10\rb\times\SI{e33}{\erg\per\second}$ for $\tpte=\lb1,3,10\rb$, meaning the $\tpte=10$ model overproduces compared to the observed values of $\SI{2.4e33}{\erg\per\second}$ \citep{Baganoff2003} and $\SI{3.6e33}{\erg\per\second}$ \citep{Nowak2012}. At $\tpte=10$ bremsstrahlung dominates near $\nu\sim\SI{e20}{\hertz}$, and at $\ms{T}{p}/\ms{T}{e}=1$ double-scattered synchrotron photons dominate. In all models, synchrotron radiation dominates the total luminosity, while bremsstrahlung amounts to a $\lesssim1\%$ contribution.

\section{Conclusion}\label{sec:conclusion}
We used the radiative transfer code \code{grmonty}, modified with a numerical bremsstrahlung prescription, to study bremsstrahlung in slowly accreting black holes as a function of accretion rate and spin. We found that bremsstrahlung is relatively constant in time and follows the expected scaling $\av{\Lum{br}}{t}\propto\av{\mdot}{t}^{2}$ in the low-accretion rate regime. Bremsstrahlung contributes considerably to the SED near $\nu\sim\SI{e20}{\hertz}$ at all accretion rates, and to the total luminosity at $\mdot\approx10^{-5}$. For \mes{}, we found that the $m=3\times10^{9}$ and $\spin=0.5$ model overproduces the observed 2--\SI{10}{\kilo\electronvolt} luminosity, and that \xray{} variability might provide information about black hole spin, although 3D models are needed to fully explore this possibility.

This work is limited by axisymmetry of the \radgrmhd{} models and by the neglect of bremsstrahlung in the \radgrmhd{} evolutions themselves; in this sense the \radgrmhd{} calculation is not self-consistent. Possible future work includes (i) \deleted{use of }3D \radgrmhd{} models that include bremsstrahlung self-consistently, (ii) \replaced{considering}{using} a non-thermal distribution of electrons, which is almost certainly required to produce the near-infrared flares in \sgra{}, and (iii) considering larger models and longer runs to model the emission from larger radii. While both bremsstrahlung and inverse Compton scattering produce photons with energy above the electron-positron pair production threshold, the optical depth to pair production remains $\lesssim10^{-3}$ in all of our models.

\acknowledgments
We thank John Wardle for comments. We also thank the anonymous referee for a report that greatly improved the presentation of the paper. This work was supported by National Science Foundation (NSF) grants AST-1716327 and OISE-1743747. RY was supported by NSF grants AST-1333612, AST-1716327, and OISE-1743747, as well as by a tuition and fee waiver from the University of Illinois at Urbana--Champaign. GNW was supported by NSF grant AST-1716327 and by the US Department of Energy through Los Alamos National Laboratory. Los Alamos National Laboratory is operated by Triad National Security, LLC, for the National Nuclear Security Administration of the US Department of Energy (Contract No.~89233218CNA000001). This work has been assigned a document release number LA-UR-19-32704.

\added{\software{\code{ebhlight} \citep{Ryan2017}, GNU Scientific Library, \code{grmonty} \citep{Dolence2009}, h5py, HDF5, Mathematica, Matplotlib \citep{matplotlib}, NumPy.}}

\appendix\label{sec:appendix}
In this section, we compare several \replaced{different formulations of bremsstrahlung emission coefficients}{bremsstrahlung emission coefficient formul\ae{}} found in the literature. In general, the \replaced{emissivity}{emission coefficient} is not analytic and must be evaluated numerically. All formul\ae{} are simplified to the case of an ionized hydrogen plasma. The latter simplification might not be valid for \sgra{}, which is likely fueled by helium-rich stellar winds \citep{Martins2007}.

\section{Electron-ion emission coefficient}\label{sec:appendix:ei}
We consider four approximations for the electron-ion bremsstrahlung emission coefficient. The first three are \replaced{based on the emissivity}{of the} form \cite[e.g.,][5.14a]{RL1979}
\begin{equation}
\jnu{br,ei}=\frac{8 \ms{q}{e}^{6}}{3\ms{m}{e}^{2}c^{4}}\sqrt{\frac{2\pi}{3}}\thte^{-1/2} n_{e} n_i e^{-h\nu/\ms{k}{B}\ms{T}{e}}\gaunt{ei},
\label{eq:EmissEIRL}
\end{equation}
where $\ms{q}{e}\equiv$ electron charge, $\ms{m}{e}\equiv$ electron mass, $c\equiv$ speed of light, $\ms{k}{B}\equiv$ Boltzmann constant, $\ms{T}{e}\equiv$ electron temperature, $\thte\equiv\ms{k}{B}\ms{T}{e}/\ms{m}{e}c^{2}$, $\ms{n}{e}\equiv$ electron number density, and $h\equiv$ Planck's constant. The thermally averaged electron-ion Gaunt factor is $\gaunt{ei}$. The three approximations based on this form of the emission coefficient differ in their treatment of $\gaunt{ei}$.

In the first approximation, the Gaunt factor is constant:
\begin{equation}
\label{eq:gffEiConstant}
\gaunt{ei}=1.2.
\end{equation}

In the second approximation (``RL piecewise''), the Gaunt factor is given by a piecewise function \citep{Novikov1973}, which for $\ms{T}{e}\gtrsim\SI{e5}{\kelvin}$ is
\begin{equation}
\label{eq:pwgaunt}
\gaunt{ei}=\begin{dcases}
\lp\frac{3}{\pi}\invu\rp^{1/2}
& \invu< 1 \\
\frac{\sqrt{3}}{\pi}\ln\lp\frac{4}{\xi}\invu\rp
& \invu>1
\end{dcases}.
\end{equation}
\added{Here $\xi\equiv\exp\lp\gamma_\text{E}\rp\approx1.781$, where $\gamma_{E}$ is the Euler-Mascheroni constant.}

In the third approximation (``RL van Hoof''), the Gaunt factor is \replaced{ interpolated from a table provided by \cite{VanHoof2015}}{computed numerically \citep{VanHoof2015}}.

The fourth, distinct approximation (``Svensson EI'') begins with a piecewise bremsstrahlung cooling rate \citep{Svensson1982}, with small corrections \citep{Narayan1995} to ensure continuity across $\thte = 1$. An approximate \replaced{emissivity}{emission coefficient} follows from multiplying the cooling rate by $\gaunt{ei}e^{-h\nu/\ms{k}{B}\ms{T}{e}}h/ \lp4\pi\ms{k}{B}\ms{T}{e}\rp$ \citep[e.g.,][]{Straub2012}:
\begin{equation}
\label{eq:EmissEIStraub}
\jnu{br,ei}=\frac{4\pi\ms{q}{e}^{6}}{3\ms{m}{e}^{2}c^{4}}\F{ei}\thte^{-1}\ms{n}{e}\ms{n}{i}e^{-h\nu/\ms{k}{B}\ms{T}{e}}\gaunt{ei},\quad\quad\F{ei}=\begin{dcases}4\lp\frac{2\thte}{\pi^{3}}\rp^{1/2}\lp1+1.781\thte^{1.34}\rp& \thte< 1 \\
\frac{9\thte}{2\pi}\ls\ln\lp 2\eta\thte+0.48\rp+1.5 \rs & \thte>1
\end{dcases},
\end{equation}
where $\eta\equiv\exp{\lp-\ms{\gamma}{E}\rp}\approx0.561$.

\section{Electron-electron emission coefficient}
A \replaced{second}{first} approximation (``Svensson EE'') is obtained using the same procedure to convert an electron-electron cooling \replaced{function}{rate} from \cite{Svensson1982} to an \replaced{emissivity}{emission coefficient} as in \ref{eq:EmissEIStraub}, yielding
\begin{equation}
\jnu{br,ee}=\frac{\ms{q}{e}^{6}}{2\ms{m}{e}^{2}c^{4}}\F{ee}\thte^{-1}\ms{n}{e}^{2}e^{-h\nu/\ms{k}{B}\ms{T}{e}}\gaunt{ee}
\label{eq:EmissEEStraub},\quad\quad
\F{ee}=\begin{dcases}
\begin{aligned}
&\frac{20\lp 44-3\pi^{2}\rp}{9\pi^{1/2}}\thte^{3/2}\\
& \times \lp 1+1.1\thte +\thte ^{2}-1.25\thte^{5/2}\rp
\end{aligned}
& \thte< 1 \\
24\thte\ls \ln\lp2\eta\thte\rp+1.28 \rs & \thte>1
\end{dcases}.
\end{equation}
\added{Similarly, \gaunt{ee} is the thermally-averaged electron-electron Gaunt factor.}
\replaced{A first approximation (``Nozawa'') has an emissivity of the form \citep{Nozawa2009}}{Another form of the emission coefficient \citep[``Nozawa,''][]{Nozawa2009} is}
\begin{equation}
\jnu{br,ee}=\frac{4\pi\ms{q}{e}^{6}}{3\ms{m}{e}^{2}c^{4}}\thte^{1/2}\ms{n}{e}^{2}e^{-h\nu/\ms{k}{B}\ms{T}{e}}\gaunt{ee},
\label{eq:jnuEeNozawa}
\end{equation}
where $\gaunt{ee}$ is given by an analytic fitting formula to numerical \replaced{data}{calculations} \citep{Nozawa2009}.

\section{Total emission coefficient}
\added{An approximation for the total emission coefficient can be constructed via multiplying the cooling rate presented in equation 5.25 of \citet{RL1979} by $e^{-h\nu/\ms{k}{B}\ms{T}{e}}h/ \lp4\pi\ms{k}{B}\ms{T}{e}\rp$ and setting $\bar{g}_{\text{B}}=1.2$, yielding
\begin{equation}
\jnu{br}=\frac{8 \ms{q}{e}^{6}}{3\ms{m}{e}^{2}c^{4}}\sqrt{\frac{2\pi}{3}}\thte^{-1/2} n_{e} n_i e^{-h\nu/\ms{k}{B}\ms{T}{e}}\bar{g}_{\text{B}}\lp1+2.61\thte\rp.
\label{eq:EmissRL}
\end{equation}
The extra factor (compared to equation \ref{eq:EmissEIRL}) accounts for electron-electron bremsstrahlung and relativistic effects \citep{Novikov1973}.}
\section{Comparison}
\label{sec:appendix:comparison}
We compare several approximate bremsstrahlung emission coefficients by computing their errors with respect to a reference emission coefficient. We choose this reference to be, for electron-ion bremsstrahlung, the emission coefficient \ref{eq:EmissEIRL} with the \cite{VanHoof2015} Gaunt factor, since they compute the Gaunt factor numerically in the relativistic regime using the Born approximation and combine the results with a previous calculation in the non-relativistic regime \citep{VanHoof2014}, thus spanning a large\deleted{ area of} parameter space with high accuracy. For electron-electron bremsstrahlung, we choose the emission coefficient and Gaunt factor from \cite{Nozawa2009} as the reference calculation \replaced{since}{because, similarly,} they \replaced{use a method analogous to that of \cite{VanHoof2015}}{merge non-relativistic results with calculations that account for Coulomb corrections and relativistic effects when important}.\par

\begin{figure*}
\gridline{\fig{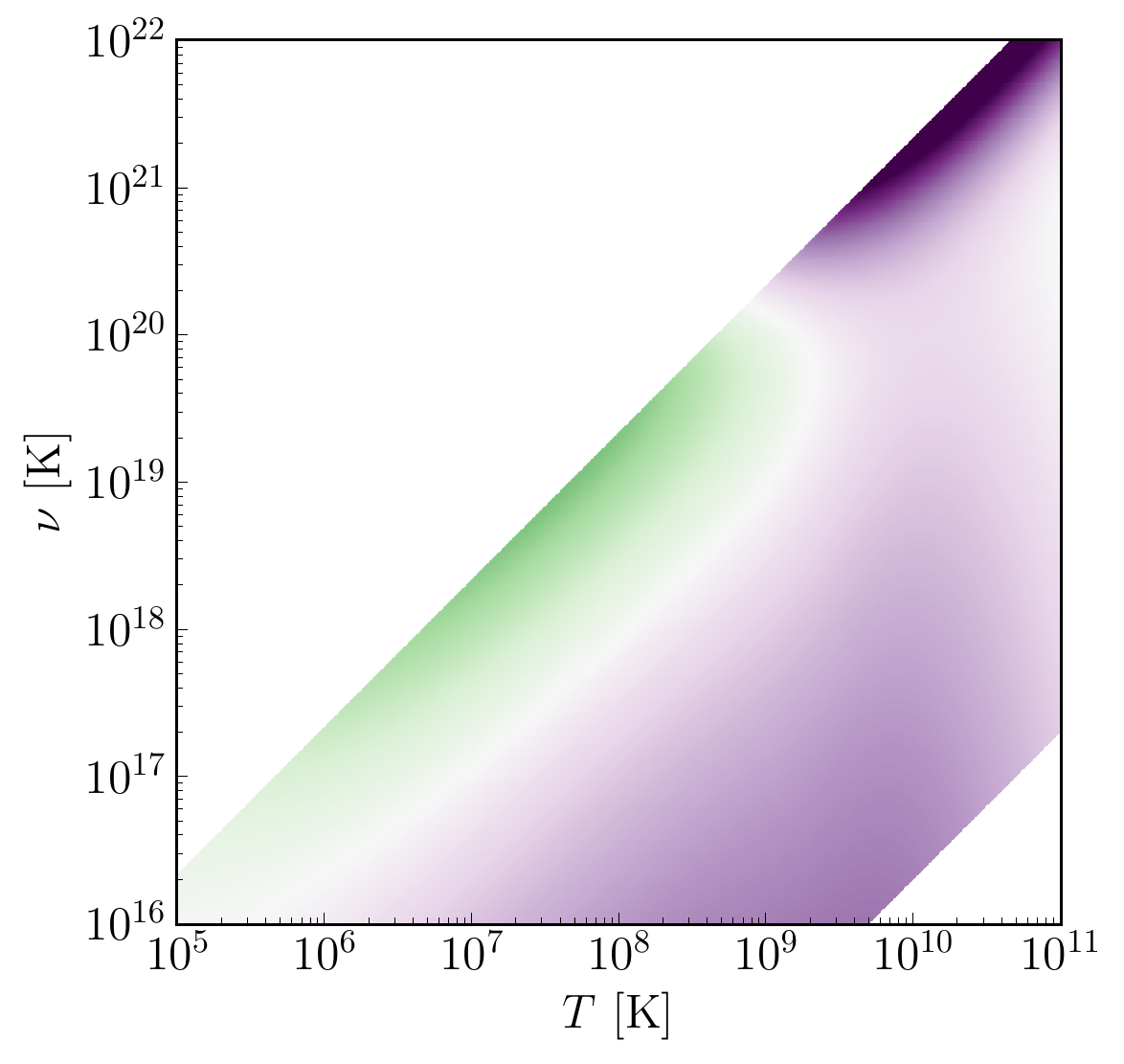}{0.458148\textwidth}{$\log_{10}\lp\frac{\displaystyle\text{\ref{eq:EmissRL} with $\gaunt{ei}=1.2$}}{\displaystyle\text{RL van Hoof}+\text{Nozawa}}\rp$}\fig{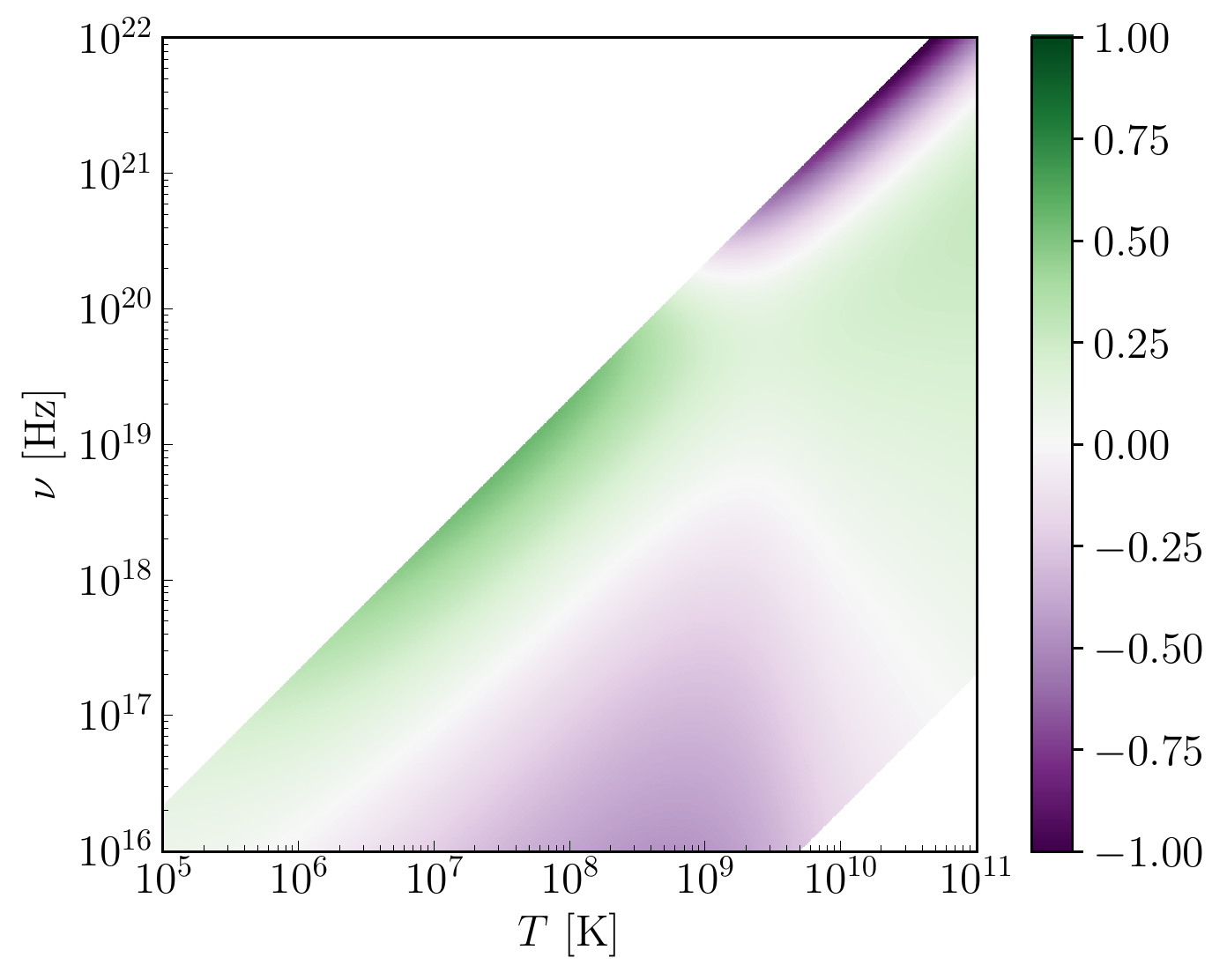}{0.541852\textwidth}{$\log_{10}\lp\frac{\displaystyle\text{Svensson EI}+\text{Svensson EE, with $\gaunt{ei}=\gaunt{ee}=1.2$}}{\displaystyle\text{RL van Hoof}+\text{Nozawa}}\rp$}}
\caption{Error for some approximations of the emission coefficient when compared to the reference calculation, taken to be the sum of equation \ref{eq:EmissEIRL} with the \cite{VanHoof2015} Gaunt factor (electron-ion contribution) and equation \ref{eq:jnuEeNozawa} with the \cite{Nozawa2009} Gaunt factor (electron-electron contribution). The left panel shows the error for \replaced{neglecting electron-electron bremsstrahlung and using a constant Gaunt factor in equation \ref{eq:EmissEIRL}}{equation \ref{eq:EmissRL} using $\bar{g}_{\text{B}}=1.2$}. The right panel shows the error for equations \ref{eq:EmissEIStraub} for electron-ion bremsstrahlung and \ref{eq:EmissEEStraub} for electron-electron bremsstrahlung, and $\gaunt{ei}=\gaunt{ee}=1.2$. \added{The shape of the plotted region reflects the domain of validity of the reference calculation.}}
\label{fig:PrescriptionComparison}
\end{figure*}

\added{We find that equation \ref{eq:EmissEIRL} using equation \ref{eq:gffEiConstant} underestimates the emission by more than an order of magnitude when $\thte>1$ and $\nu>\SI{e16}{\hertz}$, confirming the importance of electron-electron bremsstrahlung and relativistic effects in this regime. We therefore focus on formul\ae{} that include both of these. }To \replaced{do the comparison}{compare these fomul\ae{}} in the most application-agnostic way possible, we directly plot their errors as a function of frequency and temperature (Figure \ref{fig:PrescriptionComparison}). The left panel shows the error for \replaced{including only electron-ion bremsstrahlung \citep{RL1979} with a constant Gaunt factor of 1.2. The approximation does reasonably well in the low-temperature regime, but by $T\sim\SI{5e9}{\kelvin}$ (or $\thte\sim1$) it underestimates the emission by an order of magnitude (and more as the temperature increases). This resulted is expected because at these temperatures both electron-electron bremsstrahlung and relativistic corrections are important}{equation \ref{eq:EmissRL} with the constant Gaunt factor \ref{eq:gffEiConstant}. This approximation is within one order of magnitude of the reference calculation everywhere except at very high temperatures and frequencies. However, it consistently underestimates the emission by a factor of $\sim2$. The right panel shows the error for the sum of equations \ref{eq:EmissEIStraub} and \ref{eq:EmissEEStraub}, with $\gaunt{ei}=\gaunt{ee}=1.2$. This approximation has a smaller mean error and stays within one order of magnitude of the reference calculation everywhere. For these approximations, a constant Gaunt factor gives slightly better results than a piecewise Gaunt factor.}

\deleted{In sum, these approximations to the bremsstrahlung emissivity typically differ by a factor of $\sim 2$ and sometimes by more than an order of magnitude. }For applications to black hole accretion in which \replaced{this}{bremsstrahlung formul\ae{}} are the leading source of error (and we do not know of any) we recommend the RL van Hoof + Nozawa \replaced{model}{formul\ae{}}, although \added{they are more computationally expensive and} it would then be necessary to also account for helium abundance and metallicity, which we have not done here.

\bibliography{main}{}
\bibliographystyle{aasjournal.bst}
\end{document}